\documentclass{article}
\usepackage{spconf,amsmath,graphicx}
\usepackage{color}
\usepackage{subfigure}
\usepackage{bm}
\usepackage{graphicx}
\usepackage{url}

\title{F0-consistent many-to-many non-parallel voice conversion via conditional autoencoder}
\name{Kaizhi Qian$^1$\sthanks{This work was partially performed while interning at Adobe Research.}, Zeyu Jin$^2$, Mark Hasegawa-Johnson$^1$, Gautham J. Mysore$^2$\thanks{This work was funded by NSF IIS 19-10319.}}
\address{
  $^1$University of Illinois at Urbana-Champaign, IL, USA\\
  $^2$Adobe Research, CA, USA}

\newcommand{\bslname}{\textsc{AutoVC }}
\newcommand{\bslnamens}{\textsc{AutoVC}}

\begin{document}
\ninept
\maketitle
\begin{abstract}
Non-parallel many-to-many voice conversion remains an interesting but challenging speech processing task. Many style-transfer-inspired methods such as generative adversarial networks (GANs) and variational autoencoders (VAEs) have been proposed. Recently, \bslnamens, a conditional autoencoders (CAEs) based method achieved state-of-the-art results by disentangling the speaker identity and speech content using information-constraining bottlenecks, and it achieves zero-shot conversion by swapping in a different speaker's identity embedding to synthesize a new voice. However, we found that while speaker identity is disentangled from speech content, a significant amount of prosodic information, such as source F0, leaks through the bottleneck, causing target F0 to fluctuate unnaturally. Furthermore, \bslname has no control of the converted F0 and thus unsuitable for many applications. In the paper, we modified and improved autoencoder-based voice conversion to disentangle content, F0, and speaker identity at the same time. Therefore, we can control the F0 contour, generate speech with F0 consistent with the target speaker, and significantly improve quality and similarity. We support our improvement through quantitative and qualitative analysis.  

\end{abstract}
\section{Introduction}
\vspace{-0.04in}
\label{sec:intro}
\vspace{-0.02in}

Voice conversion is the process that transforms the speech of a speaker (source) to sound like a different speaker (target) without altering the linguistic content. It is a key component to many applications such as speech synthesis, animation production, and identity protection.
Conventional methods explicitly express a conversion function using a statistical model that transforms the acoustic feature (such as MFCC) of the source speaker to that of a target speaker \cite{stylianou1998continuous,kain1998spectral,toda2005spectral}. Constrained by the simplicity of the model and the vocoding algorithm that converts acoustic features to a waveform, such methods tend to produce robotic-sounding results. Recent work uses deep neural networks to address these constraints: feed-forward neural networks (DNN) \cite{chen2014voice,mohammadi2014voice} and recurrent neural networks (RNNs) such as long-short-term memory (LSTM) have been employed to replace the conversion function \cite{sun2015voice,oyamada2017non}. With the introduction of WaveNet \cite{van2016wavenet}, a host of new methods \cite{kobayashi2017statistical,liu2018wavenet,chen2018high} employed it as vocoder and vastly improved synthesis quality. 
However, most advances are in the parallel voice conversion paradigm, where parallel data (source and target speakers reading the same sentences) is required. It is in recent years that non-parallel voice conversion started gaining attention \cite{kameoka2018acvae,kameoka2018stargan,saito2018non}. In this paradigm, voice samples of multiple speakers are supplied, but the samples are not of the same sentences. It is also desirable that the voice conversion can generalize to many voices in the dataset, or even outside the dataset. Such voice conversion methods are referred to as one-to-many or many-to-many voice conversion \cite{saito2011one,ohtani2010non}. The most challenging form of this problem is called zero-shot voice conversion \cite{qian2019autovc}, which converts on-the-fly from and to unseen speakers based on only a descriptor vector for each target speaker, and possibly without any unprocessed audio examples.

Inspired by the ideas of image style transfer in computer vision, methods such as VAEs \cite{hsu2016voice,huang2018voice,kameoka2018acvae}, GANs \cite{kameoka2018stargan,fang2018high,gao2018voice} and their variants have gained popularity in voice conversion \cite{hsu2017voice,chou2018multi}. However, VAEs suffers from over-smoothing. GAN-based methods address this problem by using a discriminator that amplifies this artifact in the loss function. However, such methods are very hard to train, and the discriminator's discernment may not correspond well to human auditory perception. Moreover, the sound quality degrades as more speakers are trained simultaneously. 
There is another track of research \cite{xie2016kl,saito2018non} that uses automatic speech recognition (ASR) systems to extract the linguistic contents of the source speech and then synthesizes the target speech using the target speaker's voice. This type of method produces relatively high-quality speech but they rely on the performance of pre-trained ASRs, which again require transcribed data.

Recently, \bslnamens, a conditional autoencoder (CAE) based method \cite{qian2019autovc}, applies a simple vanilla autoencoder with a properly tuned information-constraining bottleneck to force disentanglement between the linguistic content and the speaker identity by training only on self-reconstruction. The \bslname is conditioned on a learned speaker identity embedding of the source and target speakers, making it generalizable to unseen speakers. This method also assumes that the prosodic information is properly disentangled, meaning it is either part of the speaker identity or part of the speech content. However, we found that the prosodic information appears to be partially contained in both parts, causing the F0 to flip between the source F0 contour and the F0 contour following the target voice's prosody. It is especially noticeable in cross-gender conversion where F0 changes suddenly between different genders. We hypothesize two causes for this problem: first, modeling prosody requires a substantial amount of data but the speaker embedding learned from speaker identification only observes a limited amount of samples. With insufficient information about the target speaker's prosodic pattern from the speaker embedding, the decoder is unable to generate natural-sounding F0. Second, because prosodic information is incomplete, to optimize self-reconstruction, a substantial amount of F0 information will be encoded in the bottleneck and carried over to the decoder. During voice conversion, this information conflicts with the speaker embedding resulting in F0 flipping between the source and the target. 

Therefore, we address these problems by disentangling both the speaker identity and the prosodic pattern (F0) from the speech by conditioning the decoder on per-frame F0 contour extracted from the source speaker. This modification not only ensures no source speaker F0 information leaks through the bottleneck but also makes F0 controllable via modification of the conditioned F0, which could open a new path towards deep-learning based F0 modification. Our quantitative study shows that our proposed method effectively disentangles the F0 information from the input speech signal by training on self-reconstruction with a properly-tuned bottleneck. We also compare our method to \bslname in which our converted speech has F0s significantly more consistent with the F0 distribution of the target speaker, than that of \bslname. Finally, we conducted a human listening study that shows our method improves not only F0 consistency but also sound quality and similarity from \bslname in MOS and pair-comparison studies. The remainder of the paper is organized as follows. Section~\ref{sec:method} reviews the framework and the conversion process of our system. Section~\ref{sec:exper} presents and discusses the experimental results. Section~\ref{sec:conclu} concludes the paper.

\vspace{-0.04in}
\section{Methods}
\vspace{-0.05in}
\label{sec:method}
\subsection{\bslname}
\begin{figure}[tb!]
    \centering
    \subfigure[Encoder]{\includegraphics[width=0.72\linewidth]{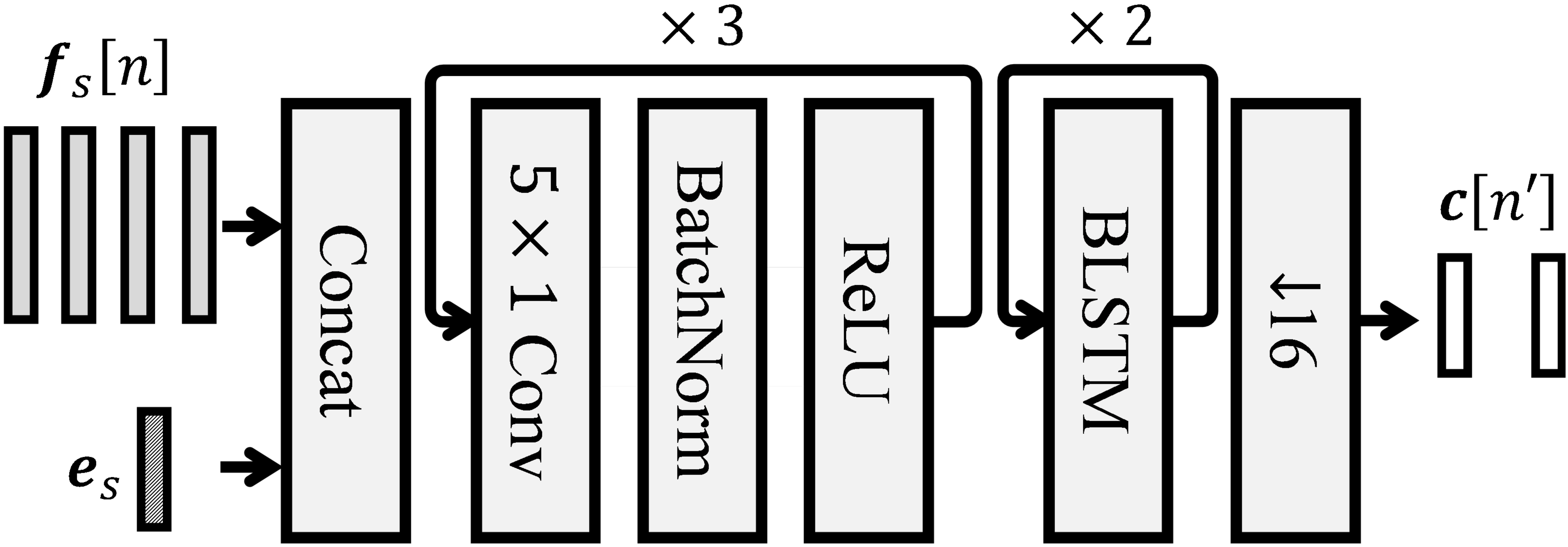}}
    \subfigure[Decoder]{\includegraphics[width=0.75\linewidth]{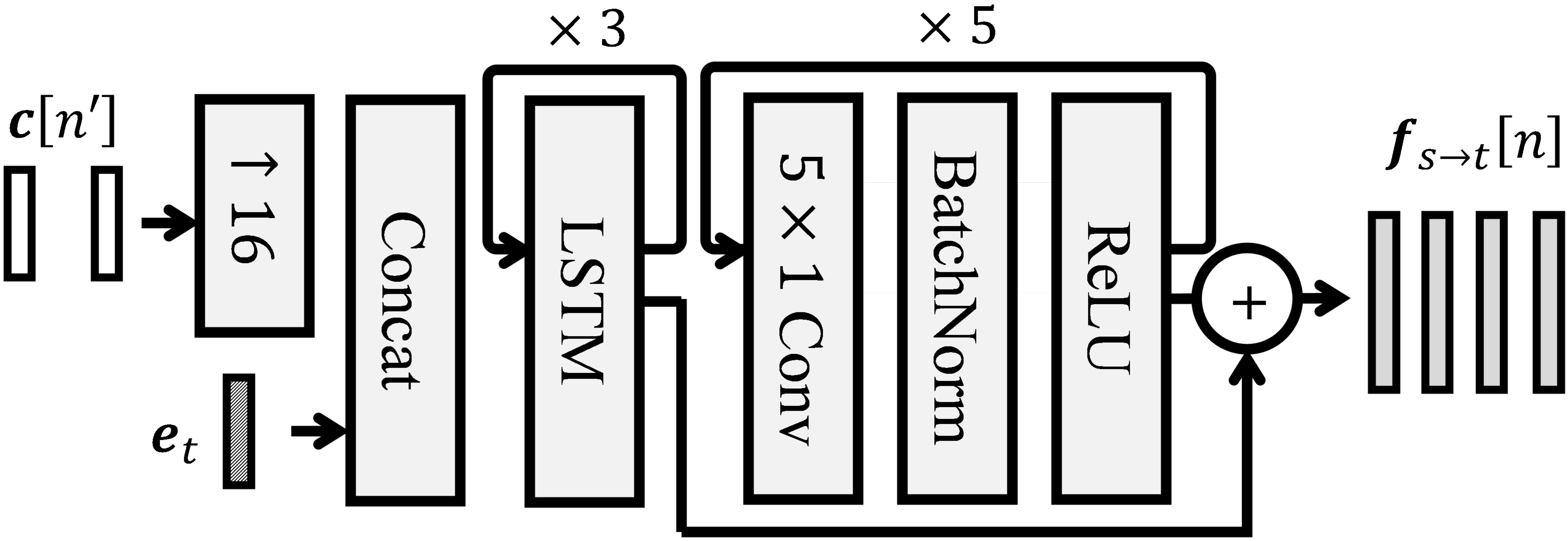}}
    \caption{The \bslname architecture. Down and up arrows denote down-sampling and up-sampling respectively. Circle arrows with `$\times n$' above denote that the enclosed blocks are repeated by $n$ times and stacked. `Concat' denotes concatenation. $\bm e_s$ and $\bm e_t$ are first copied across the time dimension before concatenation.}
    \label{fig:cae}
    \vspace{-0.08in}
\end{figure}

\bslname is a zero-shot non-parallel many-to-many voice conversion model using vanilla autoencoder \cite{qian2019autovc}. According to Fig.\ref{fig:cae}, \bslname consists of an encoder and a decoder. The encoder downsamples the input mel-spectrogram and passes it through a bottleneck to produce a content code $\bm f_s[n]$ conditioned on source speaker embedding $\bm e_s$:
\begin{equation}
    \bm c[n'] = E(\bm f_s[n], \bm e_s)
\end{equation}
where $\bm c[n']$ denotes the content code. Because sample rate changes, we use $n'$ for the indices of the code instead of $n$.
Then, the decoder takes the content code $\bm c[n]$ and synthesize the mel-spectrogram according to the target speaker's embedding $\bm e_t$ at the original sample rate:
\begin{equation}
    \vspace{-0.012in}
    \bm f_{s \rightarrow t}[n] = D(\bm c[n'], \bm e_t)
    \vspace{-0.01in}
\end{equation}
As described in \cite{qian2019autovc}, \bslname is trained on \textit{self-reconstruction only}. More specifically, during training, instead of feeding the target speaker embedding $\bm e_t$ to the decoder, we feed the source speaker embedding $\bm e_s$, leading to the self-reconstruction result, which we denote as $\bm f_{s \rightarrow s}[n]$. The training loss measures the $\ell_2$ norm of the reconstruction error in both the reconstructed speech feature and the content code, \emph{i.e.}
\begin{equation}
    \vspace{-0.005in}
    \mathcal{L} = \sum_n \lVert \bm f_{s \rightarrow s}[n] - \bm f_s[n] \rVert_2^2 + \lambda \lVert E(\bm f_{s \rightarrow s}[n], \bm e_s) - \bm c[n] \rVert_1
    \label{eq:recon_loss}
    \vspace{-0.02in}
\end{equation}
where $\lambda$ is a tunable hyperparameter. As shown in \cite{qian2019autovc}, if $\bm c[n']$ has a proper dimension and is properly downsampled, and given some other assumptions, \bslname can achieve ``perfect conversion'', in the sense that the conversion output would match the true distribution of the target speaker uttering the source content. This is because the narrow bottleneck can squeeze out the source speaker information and keep the content information only, forcing \emph{disentanglement} between speaker and content information.

Fig.\ref{fig:cae} illustrates the architecture of the encoder and decoder networks. In the encoder, the input mel-spectrogram $\bm f_s[n]$ (of dimension $80$) concatenated with the source speaker embedding $\bm e_s$ (one-hot or D-vector) at each time step is passed through three $5\times 1$ convolution layers with ReLU activation and $512$ channels, each followed by batch normalization, and then through two bidirectional LSTM layers with tunable cell dimension $16$. Finally, the resulting code is down-sampled every $16$ time steps. The down-sampling and up-sampling are different between the forward and backward outputs of the Bidirectional LSTM, as illustrated in \cite{qian2019autovc}.

In the decoder, the content code is first up-sampled by $16$, then concatenated with the target speaker embedding $\bm e_t$ (one-hot or D-vector) at each time, which is passed through three LSTM layers with cell dimension 512. Post-nets are added on top of the LSTM to refine the output mel-spectrogram\cite{shen2018natural} which consists of five $5 \times 1$ convolution layers with $512$ channels, ReLU activation except for the last layer, and batch normalization. The input to the post-net is merged to the output of the post-net through addition. The reconstruction error in the first term of Eq.~\eqref{eq:recon_loss} is evaluated both before and after the post-net.  

\vspace{-0.04in}
\subsection{F0-conditioned \bslname}
\begin{figure}
    \centering
    \includegraphics[width=0.8\columnwidth]{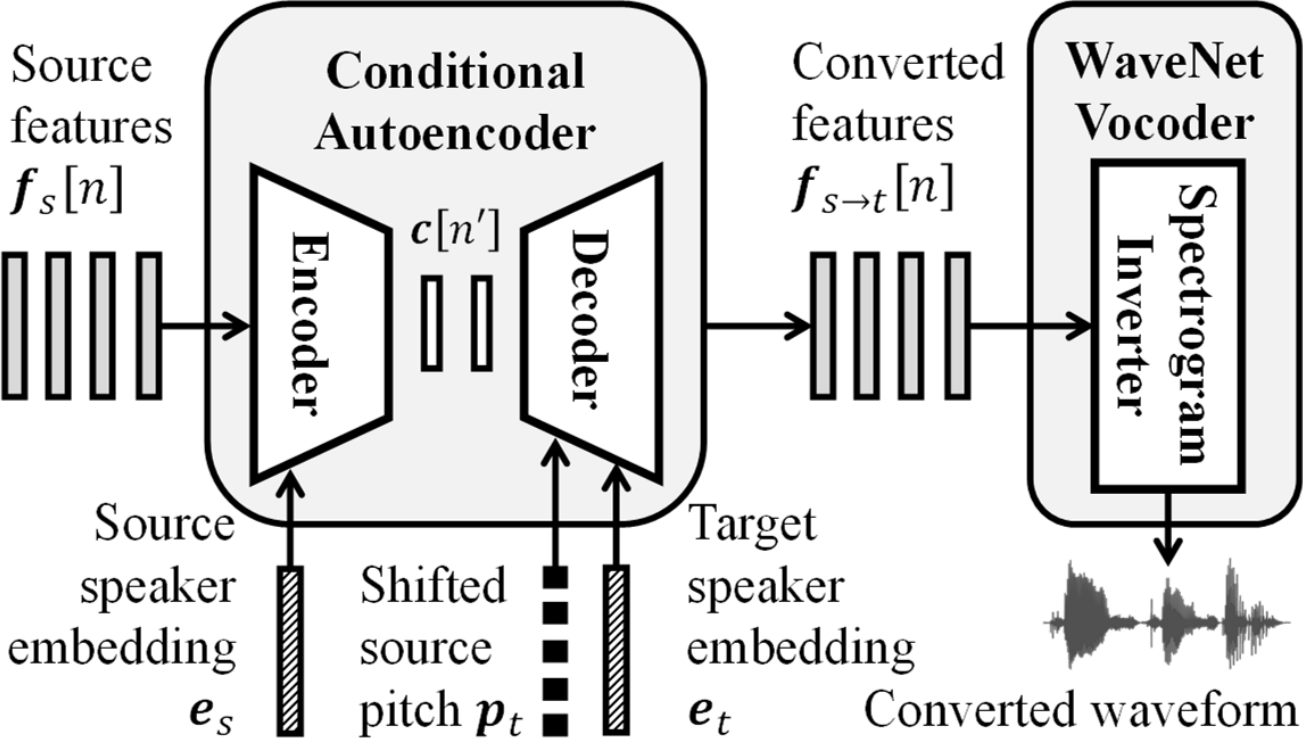}
    \caption{System Overview. The speaker embeddings are generated from waveforms by a pre-trained speaker encoder module, which is not shown in the figure.}
    \label{fig:overview}
    \vspace{-0.1in}
\end{figure}

\vspace{-0.02in}
However, speech converted using the above model contains inconsistent F0 distribution compared to the true distribution of the target speaker (please refer to Section \ref{sec:exper} for details). We hypothesize that it's caused by prosodic information (mainly F0) being encoded in the bottleneck and carried over to the decoder. As a result, the decoder generates speech that has F0 flipping between the input F0 and the target speaker's F0 pattern. The core of this issue is speaker embedding containing insufficient information about the speaker's prosodic style. One way to solve this issue is to make sure speaker embedding contains a speaker's prosody information, but it is unrealistic as it requires hours of data for each speaker. Therefore, we take another approach, disentangling all three features, speech content, F0 and speaker identity during training. Our solution is simple: in addition to speaker embedding $\bm e$, we condition the decoder of \bslname on a per-frame feature $\bm p_n$ directly computed from the source speaker's F0. This feature, called normalized quantized log-F0, is computed as follows: first, we extract the log-F0 of the source speaker's voice samples using a pitch tracker and then we compute log-F0's mean $\mu$ and variance $\sigma^2$. Then we normalize the input speech's log-F0 $p_{src}$ by $p_{norm} = (p_{src} - \mu) / \sigma / 4$. This operation roughly limits $p_{norm}$ to be within the range of 0-1. Then we quantize the range 0-1 into 256 bins and use it to one-hot encode $p_{norm}$. Finally, we add another bin to represent unvoiced frames resulting in 257 one-hot encoded feature $p_n$. Consequently, the decoder is conditioned on a global feature $e$ and per-frame feature $p_n$ as shown below:
\begin{equation}
    \bm f_{s \rightarrow t}[n] = D(\bm c[n'], \bm e_t, \bm p_n)
\end{equation}
During training, the target F0 is the source F0 normalized using the F0 mean and variance of the source speaker. Since the model is only trained on self-reconstruction loss, we expect the decoder to learn to ``de-normalize'' the conditioned F0 based on speaker embedding. The decoder architecture is the same as in Fig.\ref{fig:cae}, except that the up-sampled content code is concatenated with both the speaker embedding and F0 before feeding into the decoder, as illustrated in Fig.\ref{fig:overview}.

\vspace{-0.03in}
\subsection{Bottleneck Tuning and augmentation}
\vspace{-0.02in}
\label{subsec:augment}
Similar to \bslname, we tune bottleneck to the smallest possible size to contain sufficient speech content in order to reconstruct the mel-spectrogram of the input speech almost perfectly. Through our experiment, the bottleneck is reduced to $16$ in frequency. Since the decoder has already been provided with F0 information by conditioning on $p_n$, we hypothesize that the bottleneck will only preserve speech content. To help the decoder to learn to use $p_n$ for the missing prosodic information in the bottleneck, we augmented the data by randomly time-stretch and compress mel-spectrograms between a factor of $0.7$ to $1.35$ using interpolation. We found that that the augmentation which helps the model to generalize better to different speech rate and thus recover prosodic pattern better for the given speaker. In addition, we randomly change the signal power between $10\%$ and $100\%$ of the full power to make the model robust to volume variations. The input length is also randomly cropped between $1s$ and $3s$ to make the model robust to variable-length input. 

\vspace{-0.03in}
\section{Experiments}
\vspace{-0.02in}
\label{sec:exper}
We conducted experiments to compare the F0 consistency between converted speech of our method and \bslname. We also evaluated the proposed model under different training schemes and F0 conditions to quantitatively prove that F0 information is disentangled by the bottleneck and controllable by modifying the F0 condition at the decoder. To compare speech quality and speaker similarity of our method to \bslname, a subjective study is conducted on Amazon Mechanical Turk where subjects are asked to rate the mean-opinion-score for converted audio samples. All models are trained and evaluated on the VCTK corpus \cite{veaux2016superseded}. To be consistent to previous work, we used voice samples from the same 10 male and 10 female speakers in the experiment. The utterances of each speaker are partitioned into $90\%$ training and $10\%$ test. \bslname and our proposed models are trained using Adam optimizer with a batch size of 2 for 700k iterations with data augmentation\ref{subsec:augment}. The learning rate is $0.0001$, and $\lambda = 1$. Audio samples are available at \url{https://auspicious3000.github.io/icassp-2020-demo}

\subsection{Quantitative Analysis}

\subsubsection{F0 distribution}

\begin{figure}
    \centering
    \includegraphics[width=0.8\columnwidth]{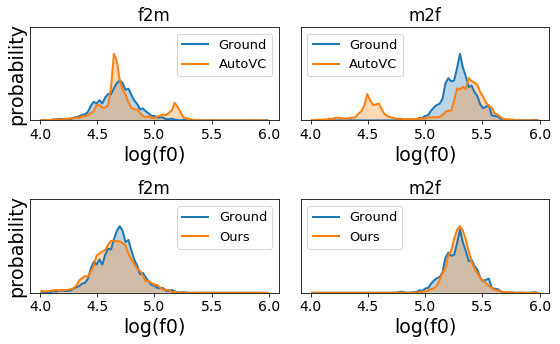}
    \vspace{-0.3cm}
    \caption{Comparison of F0 distributions between the ground truth and the generated speech using \bslname or our method.}
    \label{fig:exp1}
\end{figure}

The first study aims to illustrate the F0 issue in \bslname by comparing the F0 distribution of the converted speech with the ground truth distribution. In this study, 8 voices are used in which 4 are male and 4 are female. We computed all pairs of conversion from male to female (m2f) and from female to male (f2m), each consists of 16 pairs of voices and 10 different held-out utterances, totaling 160 samples for either case. Then we extracted the log-F0 of these samples and plot the distribution. Note that unvoiced F0s are thrown away from the plot. Finally, we overlaid the ground truth F0 distribution of the target speakers on the above two distributions, as shown in Figure~\ref{fig:exp1}.

\begin{figure*}
    \centering
    \subfigure[\bslname vs. our method]{
        \includegraphics[width=0.63\columnwidth]{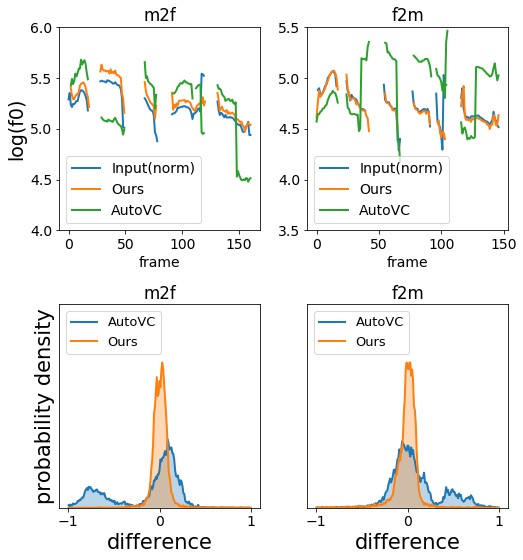}
        \label{fig:exp2}
    }
    ~
    \subfigure[our network retrained without F0 conditioning vs. our method without modification ]{
        \includegraphics[width=0.63\columnwidth]{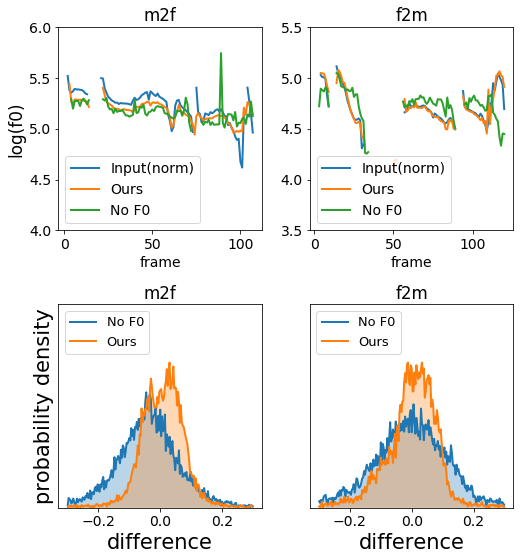}
        \label{fig:exp3}
    }
    ~
    \subfigure[Our method with the F0 condition modified to flat F0 vs. our method without modification ] {
        \includegraphics[width=0.645\columnwidth]{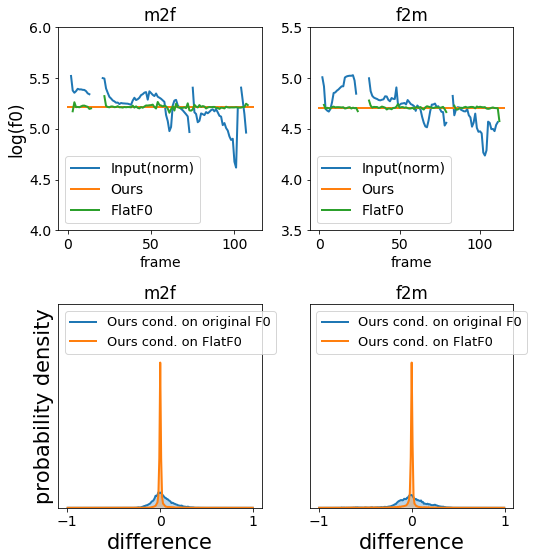}
        \label{fig:exp4}
    }
    \vspace{-0.05in}
    \caption{Comparison of F0 contours of generated speech based on the two methods. In each sub-figure, the upper two plots display example F0 contour overlaid on the input F0 normalized to the target speaker's F0 range. The lower two plots show the error distribution between the F0 of the converted speech and the normalized input F0. In both upper and lower plots, the left one corresponds to male-to-female cases and the right one corresponds to female-to-male cases. The caption of each sub-figure shows the two methods being compared.}
    \vspace{-0.1in}
\end{figure*}

From the result we can see that in both f2m and m2f cases, \bslname has two peaks in the distribution where one of them overlaps with the ground truth distribution and the other is centered at the F0 of a different gender. This is consistent with the ``flipping F0 issue'' we discussed earlier. It is also visible that m2f contains more ``F0 flipping'' than f2m. In contrast, the proposed method produces an F0 distribution that overlaps well with the ground truth distribution even though we did not explicitly tell the decoder the F0 range of the target speaker. Our hypothesis is that by conditioning on normalized F0, the target F0 range is inferred from the speaker embedding. As expected, the F0 of the output matches the speaker identity and thus consistent with the speaker's true F0 distribution. 

\vspace{-0.06in}
\subsubsection{F0 consistency}
\vspace{-0.02in}
The second study aims to measure how well the generated F0 follows the input F0. Since there lacks a ground truth F0 of the target speaker, we created a pseudo-F0 by de-normalizing the conditioned F0 using the target speaker's F0 statistics. This is equivalent to computing Gaussian normalized transformation from the source log-F0 using mean and variances of source and target voices:
\begin{equation*}
    \vspace{-0.01in}
    \log{p_{tgt}} = \mu_{tgt} + \frac{\sigma_{tgt}}{\sigma_{src}} (\log{p_{src}}-\mu_{src})
    \vspace{-0.01in}
\end{equation*}
With the pseudo-F0 in the log space $\log{p_{tgt}}$, we compare how the generated speech's F0 matches the pseudo-F0 and plot the distribution of errors for both \bslname and our method in Figure \ref{fig:exp2}. The upper half of the figure shows an actual instance in which the pseudo-F0, the F0 of our method's converted speech, and that of \bslname are plotted together. One can see that our method's F0 follows the pseudo-F0 consistently with only minor shifting, which is reasonable as the network is never trained using denormalized F0, to begin with. In contrast, the F0 produced by \bslname rapidly fluctuates above and below the pseudo-F0, and it is only partially consistent in trend. We hypothesize that it is due to F0 leaking through the bottleneck during training and thus interfering with the F0 range encoded in the speaker identity. The lower half of the plot shows the error distribution between the converted speech's log-F0 and the pseudo ground truth. Our method shows significantly smaller error rate than that of the original \bslnamens.

\vspace{-0.04in}
\subsubsection{Bottleneck test and F0 controllability}
\vspace{-0.02in}

This study focuses on experimentally verifying that our model disentangles F0 by information-constraining bottleneck and thus makes F0 controllable. In the first experiment, we concatenated an encoder from a pre-trained model with a new decoder that is only conditioned on the speaker identity without the F0. Then, we train this decoder with the encoder fixed. The resulting F0 of the converted speech becomes random as depicted in \ref{fig:exp3}. Since no source F0 information leaks through the bottleneck, the generated speech matches the F0 distribution of the target speaker but sounds random and lacks details. This result verifies our assumption that the speaker embedding only encodes  prosodic pattern partially and that the source voice's F0 information is largely disentangled by the bottleneck. To test controllability, we modified the conditioned F0 to be constant-valued (Flat F0). As shown in Figure \ref{fig:exp4}, the converted speech's F0 follows a flat contour despite that the input speech has a non-flat F0. Note that there are some F0 fluctuations at boundaries between voiced and unvoiced segments, which is likely caused by inaccuracy of F0 detection algorithms. The lower half of the plot also shows high consistency between our method's F0 and the reference flat F0. 

\vspace{-0.03in}
\subsection{Qualitative Analysis}
\begin{figure}[!t]
    \centering
    \includegraphics[width=0.82\columnwidth]{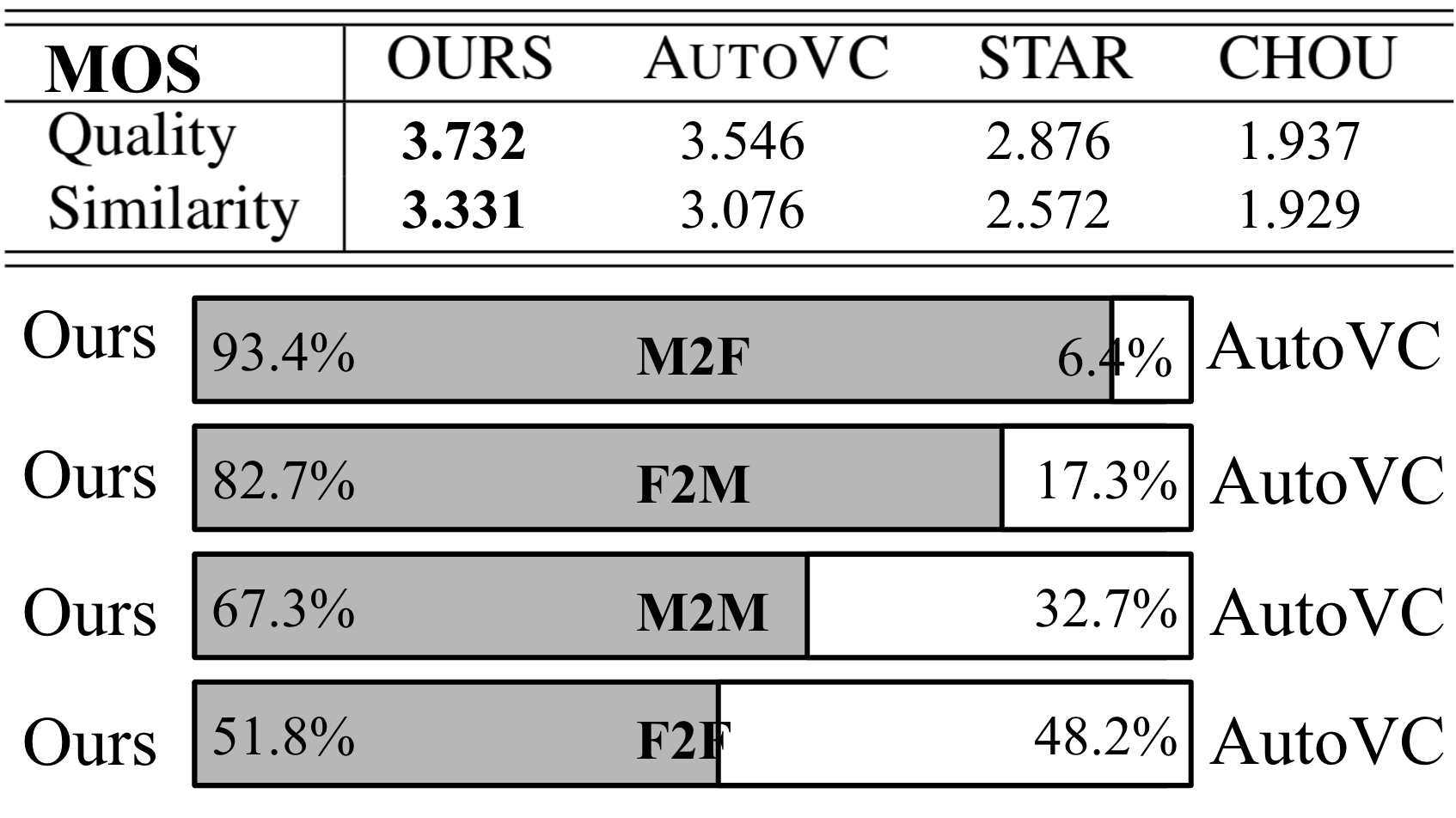}
    \vspace{-0.3cm}
    \caption{MOS and pair-comparison between \bslname and our method.}
    \label{fig:mos}
    \vspace{-0.2in}
\end{figure}

\vspace{-0.02in}
We conducted Mean-Opinion-Score (MOS) evaluation via Amazon Mechanical Turk, where subjects are asked to rate the similarity and quality of synthesized voice samples on a scale of 1-5. 
Our main goal is to compare the F0-conditioned \bslname against the original \bslname, but we also include 2 additional baselines, which we name STAR and CHOU respectively. STAR \cite{kameoka2018stargan} is a voice conversion system based on the StarGAN scheme. CHOU \cite{chou2018multi} an autoencoder based voice conversion system that adopts adversarial training to force speaker disentanglement. As shown in Figure.\ref{fig:mos}, our method with F0 disentanglement outperforms the original \bslnamens. To further verify that our method improves over \bslname in most cases, we conducted pairwise comparison tests on Turk where subjects are asked to choose between two converted speech (ours and \bslname) which one sounds better given a voice sample of the target speaker. We collected 16 ratings for each one of the 560 tests. The result is also shown in Figure \ref{fig:mos} and our method significantly outperforms the baseline especially under cross-gender conversion cases. 


\vspace{-0.06in}
\section{Conclusion}
\vspace{-0.06in}
\label{sec:conclu}
In this paper, we proposed an F0-conditioned voice conversion system that refreshes the previous state-of-the-art performance of \bslname by eliminating any F0-related artifacts. It experimentally verified the hypothesis that any conditioned prosodic features can be disentangled from the input speech signal in an unsupervised manner by properly tuning the information-constraining bottleneck of a vanilla autoencoder. This could open a new path towards more detailed voice conversion by controlling different prosodic features.

\pagebreak

\bibliographystyle{IEEEbib}
\bibliography{strings,refs}

\end{document}